# Enhanced Pulse Compression within Sign-Alternating Dispersion Waveguides

## Haider Zia[1,*]

[1]*University of Twente, Department Science & Technology, Laser Physics and Nonlinear Optics Group, MESA+ Research Institute for Nanotechnology, Enschede 7500 AE, The Netherlands*
*\*h.zia@utwente.nl*

**Abstract:** Repeatedly alternating the sign of dispersion along a waveguide substantially increases the bandwidth to input peak power efficiency of supercontinuum generation (SCG). Here, we explore, theoretically and numerically, how to optimize sign-alternating dispersion waveguides for nonlinear pulse compression, so that the compression ratio is maximized. By exploring a previously unknown SCG phase effect, we find emergent phase effects unique to these structures where the spectral phase converges to a parabolic profile independent of uncompensated higher-order dispersion. The combination of an easy to compress phase spectrum, with low input power requirements, then makes sign-alternating dispersion a scheme for high-quality nonlinear pulse compression that removes the need for high powered lasers. Also, we show a new scheme for the design of practical waveguide segments that can compress SCG pulses to near transform-limited durations, which is integral for the design of these alternated waveguides and in general, for nonlinear pulse compression experiments. We conclude by showing how compression can be maximized for alternating dispersion waveguides within the integrated photonics platform, showing compression to two optical cycles.

## 1. Introduction

Supercontinuum generation (SCG) in Kerr nonlinear waveguides is central in numerous applications such as in the generation of sub-cycle pulses [1-6], metrology using optical frequency combs [7-11], optical coherence tomography [12-14] and as a wide-bandwidth source for ranging and sensing applications [15,16]. We have recently introduced the concept of repeatedly sign-alternating dispersion along the propagation direction as a means of overcoming spectral stagnation that results in SCG waveguides [17]. Sign-alternating the dispersion maintains spectral bandwidth generation in normal dispersion segments (ND) by temporal compression of the pulse in anomalous dispersion segments (AD). Spectral generation is also kept ongoing in AD segments by the compression of the chirped pulse input from the previous ND segments, and avoiding the formation of bandwidth-stagnant solitons by renewed temporal broadening in subsequent ND segments.

While our concept of overcoming spectral stagnation enhances the input peak power to bandwidth generation efficiency significantly, we would like to explore the impact of sign-alternating dispersion for nonlinear pulse compression, both theoretically and numerically here. The motivation is that nonlinear compression could now take place at much less input peak powers, because of the more efficient bandwidth generation in the alternating structures. Lasers that otherwise have too low of a peak power, such as high-repetition sources, integrated chip lasers, or fiber oscillators could directly be used to generate ultrashort pulses without the need of multiple amplification stages. In addition, material damage that accumulates with high powered lasers would not be present, thus, a wider range of waveguide geometries and materials become accessible when considering sign-alternating dispersion waveguides for nonlinear pulse compression.

Of primary concern when considering sign-alternation for pulse compression, is the impact of a complex spectral phase profile that could develop in ND and AD segments. If this phase is present it can limit temporal compression in the AD segments, limiting the total bandwidth generation and the overall achieved temporal compression. In this paper, the impact of uncompensated higher than second-order dispersion and self-phase modulation (SPM) in the ND and AD segments will be related to the spectral phase development of the SCG pulse.

To explain the spectral phase development, we will introduce a previously unknown phase effect of SCG. This effect reduces higher-order spectral phase coefficients so that the phase remains near parabolic despite significant uncompensated higher than second-order dispersion. Given this effect, we will then show that, under certain conditions, the specific shape of the ND GVD profile does not play a significant role in determining higher-order spectral phase. Thus, a wide choice of ND segments become possible for high-quality pulse compression in the alternating structures. Of fundamental interest is that this newly discovered phase effect explains more rigorously why, in general, ND SCG has a near parabolic spectral profile, numerically or experimentally found in various experiments in the last decade [18-21].

Ultimately, we use the found spectral phase to obtain the ideal AD segment group velocity dispersion (GVD) profiles needed to make sign-alternation feasible as a method for nonlinear pulse compression. Practically obtaining the needed AD segment GVD profile then becomes the next open question we address in this paper. To address this, we propose a novel scheme to construct arbitrary AD dispersion profiles by breaking up the AD segment into sub-segments. This procedure can be used for general pulse compression experiments as well.

We then turn to a design recipe for optimum pulse compression within the integrated photonics platform, where both AD and ND segments contribute substantially to spectral generation. We simulate an example structure on the silicon nitride platform to demonstrate our approach, to obtain compression to two optical cycle pulses.

## 2.    General conditions for Segments for Nonlinear Pulse Compression

We begin by finding general conditions for the AD and ND segments in sign-alternating dispersion waveguides, that would favor high nonlinear pulse compression ratios. Of first importance is that the choice of where substantial SCG occurs (ND segment, or AD SCG ) is chosen such that higher-order spectral phase is minimized.

While normal dispersion has a saturating effect on bandwidth generation, as opposed to zero dispersion, the spectral amplitude profile contains fewer modulations, and the spectral phase profile does not have significant additional higher-order contributions from SPM. This is because normal dispersion changes the SPM induced temporal phase function in the wings of the Gaussian-like pulse, such that the phase function maintains a linear increase here as well, instead of saturating. This temporal phase effect, unique to ND SCG is called the wave-breaking effect [22] and renders that for optimal pulse compression, segments that generate spectral bandwidth should be the ND segments.

Contrary to the ND case, SPM in AD segments increases higher-order phase contributions and spectral modulations due to features that emerge in the wings of the pulse, soliton fission and modulation instability. The spectral modulations and higher-order phase render that both the transform limited duration is increased and compression to it is harder to achieve. On the other hand, the pulse duration decreases in AD SCG, so that the pulse is already in a compressed form at the output, as opposed to ND SCG where the pulse has significant second-order phase that requires further compression.

Favoring generation in ND segments is typically not problematic as ND step-index fibers and integrated waveguides exhibit a higher nonlinear coefficient from the narrower core diameters than in AD waveguides. Dominant ND SCG is then practically guaranteed if the input pulse energy remains below the fundamental soliton energy of the AD segments. The low power criterion fits nicely with the overall motivation for nonlinear compression with sign-alternating dispersion as we seek to minimize power requirements for this process. The spectral phase development in ND segments will then decide the AD GVD profiles needed.

For high power applications, the AD segment becomes active in the bandwidth generation process. Results and optimization conditions under this consideration will be described in the section 8, in the context of integrated photonics waveguides, where even with the decreased quality of compressible spectra from AD SCG two-cycle pulses were generated.

In order not to complicate the analysis we omit the Raman contribution and self-steepening to focus on SPM as the nonlinear process behind spectral generation. This omission is justified at least for self-steepening as it has a similar impact to the spectral phase development as SPM. However, the Raman contribution would ultimately impact pulse compression, and waveguides should be chosen to minimize this effect. We also assume that the waveguide structure is endlessly single moded across the frequency bandwidth range, e.g. as in the case with specialized photonic crystal fibers [23].

Ultimately, loss of single-mode propagation, Raman effects and spontaneous four-wave mixing introduces modulations and complex phase features in the spectrum that reduces the compression ratio in sign-alternating structures.

## 3. Regimes of Supercontinuum Generation within Sign-alternated Structures

To find properties of the spectral phase development of the SCG pulse, relevant to designing the corresponding pulse compression segments, we first need to define the type of spectral generation within the segments of the sign-alternated structure. The spectral phase of the SCG pulse is heavily dependent on whether self-phase modulation (SPM) dominates or dispersion. The relationship between SPM and dispersion is described by two characteristic lengths in sign-alternated waveguides, the first being the nonlinear length, $L_{nl} = \frac{1}{\gamma P}$, $\gamma$ ($\frac{1}{Wm}$), $P$ being the input peak power. The second characteristic length is the dispersion length, $L_D = \frac{\tau_o{}^2}{|d_2|}$, $\tau_o$ is the intensity 1/e half duration of the pulse entering a segment, $d_2$ is the second order GVD coefficient [19].

The ratio of these characteristic lengths, $R \equiv \frac{L_{nl}}{L_D}$, at a segment in the alternating dispersion structure dictates the segment's regime of supercontinuum generation. For $R \equiv \frac{L_{nl}}{L_D} < 1$, SCG is in the "SPM dominated regime", while for $R > 1$, SCG is in the "dispersion dominated regime".

In the SPM dominated regime, the dispersive broadening of the pulse duration occurs much more slowly as a function of propagation distance compared to the spectral bandwidth broadening. In contrast, for the dispersion dominated regime, the additional bandwidth increase from the nonlinear generation is significantly less than the original bandwidth and there is less spectral generation than within the SPM dominated regime. Here, the pulse temporally spreads in approximately the same way as the original pulse would spread due to dispersion.

$R$ decreases for higher segment numbers in the waveguide because of the additional frequency generation across prior segments, yielding shorter transform-limited pulse durations. At the end of the waveguide, the region would have likely transitioned to the dispersion dominated regime, even if it was SPM dominated at the start of the waveguide structure [17].

## 4. Convergence to Near Parabolic Spectral Phase in the SPM Dominated Regime

In this section, we show that the specific dispersion profile in the ND segment does not play a large role in determining the spectral phase, when the effects of SPM dominates SCG in the

ND segment, i.e., $R < 1$. We then show what the AD GVD profiles would be, given this robustness to higher order dispersion in the ND segments. This invariance to the dispersion profile, when SPM dominates, is the inverse analog to the optical wave-breaking effect in ND waveguides. I.e., when $R < 1$, SPM lowers higher-order phase contributions from dispersion rather than the latter lowering higher order SPM phase contributions (in the wave-breaking effect).

To understand the effects of SPM on spectral phase, we start by noting that the pulse 1/e duration, $\tau$, is slowly increasing compared to the increase of the bandwidth in the SPM dominated SCG regime. This slow increase of the pulse duration results that past a certain propagation distance, the instantaneous frequencies along the temporal frequencies start to deviate from the central pulse frequency much more rapidly than increasing their location in time due to dispersion. This steeper increase in 1/e spectral bandwidth, $\Delta\nu$, compared to duration in relation to the propagation coordinate results in the spectral phase, $\varphi$, profile being stretched over a larger bandwidth without correspondingly increasing its values to maintain its shape (see illustration in Fig. 1).

The impact of this stretching of the spectral phase results in the decrease of its curvature and higher order derivatives with respect to frequency, whereby these derivatives are progressively reduced by a factor related to the bandwidth increase along the propagation. The progressive reduction of higher-order derivatives results in a dominant parabolic profile to emerge, independent of higher order dispersion in the waveguide. In this section, this reduction will be quantitively shown, numerically and analytically, through how the relative magnitudes of Taylor coefficients of $\varphi$ taken about the central angular frequency, $\nu_o$, changes along the propagation coordinate.

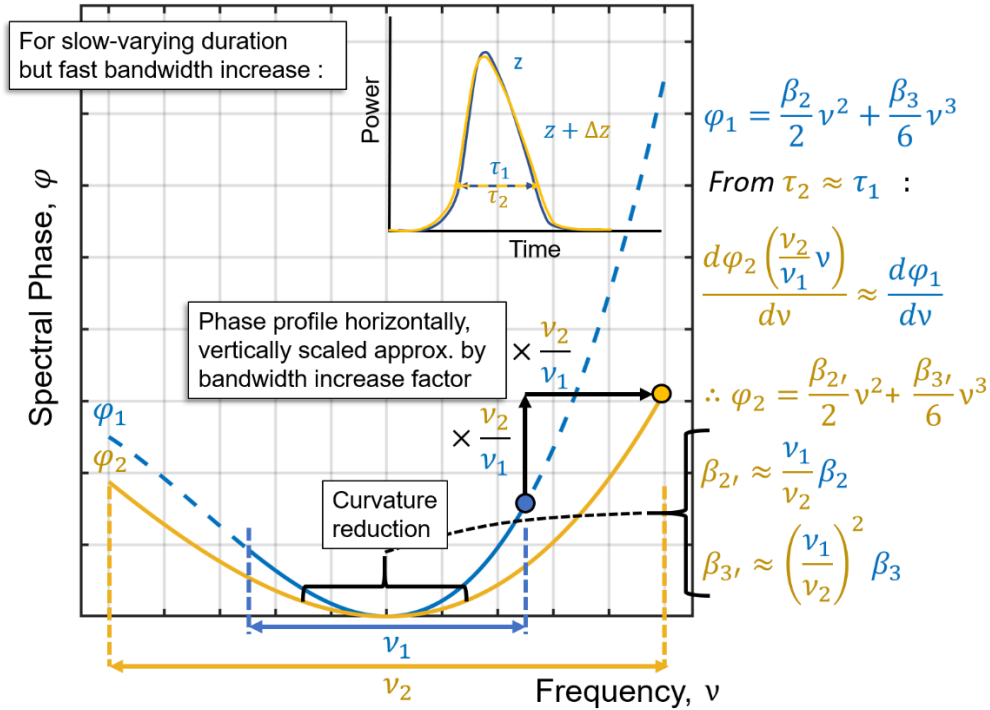

Fig 1: Illustration sketch of the spectral phase function of a pulse that undergoes bandwidth broadening while maintaining only a small increase in pulse duration (pulse power versus time shown in inset). The phase spectrum at two positions of propagation are shown, the blue spectrum being at position $z$, while the yellow spectrum is when the pulse is at a later position $z + \Delta z$. The curvature of the spectral phase function decreases, resulting in a lowering of phase

derivatives (phase-order coefficients) around the central frequency as seen in the yellow curve compared to the blue one. The spectral phase function for the yellow curve is derived from the condition that the location of instantaneous frequencies in time remains unchanged as they are increased in frequency by the new bandwidth generation. It is shown that the scaled phase function has phase order coefficients that are exponentially lowering with order number, as shown in the equations of the phase derivatives on the right-hand side. This lowering converges the phase profile to parabolic. The dotted blue curve shows the extension of the spectral phase function for the pulse at $z$ over the bandwidth range of the pulse at $z + \Delta z$ to highlight the relative flattening of the phase at this later position.

To demonstrate the spectral phase development numerically, we solve the generalized 1-D nonlinear Schrödinger equation with dispersion terms up to the 20th order. The nonlinear Schrödinger equation under the slowly varying envelope approximation [1], is given as

$$\frac{\partial u}{\partial z} = \sum_{k \geq 2} \frac{i^{k+1}}{k!} \beta_k \frac{\partial^k u(z,\tau)}{\partial \tau^k} + i\gamma \left(1 + i\tau_s \frac{\partial}{\partial \tau}\right)|u(z,\tau)|^2 u(z,\tau) \ , \tag{1}$$

where $u$ is the complex field envelope, $\beta_k$ are the Taylor series coefficients of the expansion of the frequency-dependent wave number about the central frequency, $\omega_o$. $\tau = t - V_g z$ is the time coordinate, co-moving in the frame of reference of the group velocity dispersion ($V_g \equiv \beta_1^{-1}$). $\tau_s$ is the characteristic timescale of self-steepening, which is set to zero since we omit this effect in our analysis.

The equation is evaluated using the Strang split-step exponential Fourier method [24] iteratively along the propagation coordinate $z$. The dispersive contributions are evaluated in the frequency domain using as input the frequency-dependent group velocity dispersion of the ND segment under consideration, obtained from a waveguide mode solver.

In this numerical model, we input the ND segment GVD profile of a normal dispersion fiber (Corning Hi1060flex [17]) with substantial negative third-order dispersion (TOD). We show the GVD of this fiber in Fig. 2, plotted along the envelope angular frequency range of interest. The carrier angular frequency ($193.4\,\text{THz}$), corresponding to the zero of the envelope frequency in this example, corresponds to a pulse input wavelength of $1.55\,\mu\text{m}$.

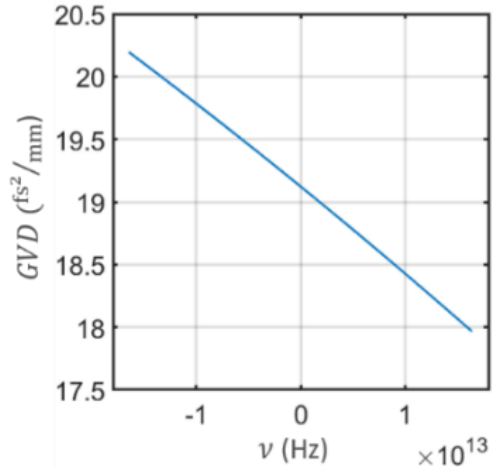

Fig. 2: Group velocity dispersion in $fs^2/mm$ plotted with respect to envelope angular frequencies, across the bandwidth range of interest in an example normal dispersion fiber (Corning hi1060flex).

For input into the numerical model a transform-limited Gaussian pulse (1/e power duration of 72 fs), with a pulse energy of 2 nJ is used. The peak power and duration of this input pulse renders that SCG here is in the SPM dominated regime of the ND fiber. With these parameters, we find that the typical characteristic spectral phase evolution of the SCG pulse in the SPM dominated regime is shown.

We plot the 1/e angular frequency bandwidth of the spectral energy density, $\Delta\nu$, against the propagation distance, $\Delta z$, within the ND fiber in Fig. 3a. The resulting saturation curve is characteristic of spectral bandwidth development in ND SCG within the SPM dominated regime. We show the propagation up to 30 cm, which is approximately the saturation length, $L_{sat}$, of the bandwidth development.

*SPM effects on second-order phase*

We start our analysis of the Taylor coefficient development of the spectral phase with the second-order contribution ($\beta_2 \equiv \frac{d^2\varphi}{d\nu^2}\big|_{\nu_o}$). We plot the ratio of $\beta_2$ with SPM to the $\beta_2$ (labelled as $\beta_{2o}$) without SPM, versus propagation in Fig. 3b, obtained from the GNLSE numerical simulation. $\beta_{2o}$ is equal to $d_2\Delta z$, $d_2$ is the waveguide GVD coefficient (taken about $\nu_o$).

Here we see qualitatively the expected trend as described above, i.e., a decrease past a certain propagation distance (i.e., when the pulse obtains a specific bandwidth value). Then when spectral development ceases, close to the saturation length, the reductive effect of SPM on $\beta_2$ stops and $\beta_2$ asymptotically approaches $\beta_{2o}$ i.e., the ratio plotted in Fig. 3b approaches one. However, at the beginning of the pulse propagation there is an increase in $\beta_2$ to a peak value at approx. 2 cm, before its decrease.

We now proceed to explain why $\beta_2$ increases in the beginning of the propagation and the conditions for its decrease after a certain propagation is reached. The increase in bandwidth as a function of propagation decreases the transform-limited duration of the pulse, $\tau_{lim} \propto \frac{1}{\Delta\nu}$. The decreasing $\tau_{lim}$ results in more spectral phase being added across the spectrum to change the temporal duration of the pulse from $\tau_{lim}$ to $\tau$, i.e., by adding $\Delta\tau = \tau - \tau_{lim}$. This causes the second order spectral phase coefficient, $\beta_2$, to be an increasing function of pulse propagation.

The increasing $\beta_2$, however goes to a peak value, then decreases. This is explained by two processes: Firstly as $\tau_{lim}$ asymptotically approaches zero due to the bandwidth generation, $\Delta\tau$ approaches $\tau$, i.e., $\Delta\tau \to \tau$. Since, the pulse duration increases slowly compared to the rate of spectral bandwidth generation as a function of propagation distance $\beta_2$ increases slowly as a function of $\Delta\tau$, in this asymptotic limit of $\tau_{lim}$.

Secondly, on the other hand, even while $\Delta\tau$ becomes slowly increasing, the frequency bandwidth still rapidly increases. Increasing the bandwidth while only marginally increasing the pulse duration, results in a decreasing $\beta_2$ when $\tau_{lim} \to 0$, since the (half-)duration of the pulse is simply the time location of the 1/e instantaneous angular frequency in relation to the center of the pulse. I.e., $\tau = 2\frac{d\varphi}{d\nu} \approx 2\beta_2\Delta\nu$ (when the pulse center is considered zero time). Since $\tau$ is slowly varying in relation to $\Delta\nu$ it can be considered a constant, therefore, by rearranging for $\beta_2$, it is found that $\beta_2 \propto 1/\Delta\nu$ and decreases as a function of pulse propagation.

Given the above discussion, we derive the following conditions from the relationship between the second-order spectral phase, spectral bandwidth, and the corresponding Gaussian pulse duration. It can be shown that $\beta_2$ reaches its peak when $\tau \approx \sqrt{2}/\Delta\nu$, occurs at a certain propagation distance within the ND waveguide. After which $\tau > \sqrt{2}/\Delta\nu$ and $\beta_2$ will be decreasing if still in the SPM dominated regime of SCG. When this equality is satisfied, by using the Taylor series representation of the spectral phase, an upper bound scaling of $\beta_2(\Delta z)$ in relation to the bandwidth can be found:

$$\tau = 2 \frac{d\varphi}{d\nu}\Big|_{\frac{\Delta\nu}{2}} = \sum_{n=2}^{\infty} \frac{2}{n-1!} \frac{d^n\varphi}{d\nu^n}\Big|_{\nu_o} \left(\frac{\Delta\nu}{2}\right)^{n-1} \tag{2}$$

Rearranging for $\beta_2$, we arrive at:

$$\beta_2 = \frac{\tau}{\Delta\nu} - \sum_{n=3}^{\infty} \frac{1}{n-1!} \frac{d^n\varphi}{d\nu^n}\Big|_{\nu_o} \left(\frac{\Delta\nu}{2}\right)^{n-1} \tag{3}$$

Eq. 3 is always greater than zero, therefore, at most $\beta_2$ scales as

$$\beta_2(\Delta z) \propto \frac{\tau(\Delta z)}{\Delta\nu(\Delta z)}, \tag{4}$$

as a function of propagation in the waveguide.

Going back to Fig. 3b, for most of its development, $\beta_2$, is higher than $\beta_{2o}$, and only dips to 96% the value of $\beta_{2o}$. A stronger reduction of $\beta_2$ below $\beta_{2o}$ can be achieved, for example, with input pulses at shorter pulse durations where the peak power is raised such that the $R$ ratio is conserved. However, sufficient for pulse compression is to maintain a parabolic profile despite higher-order uncompensated dispersion. Given this sufficiency, the criterion simply becomes that regardless of the value of $\beta_2$, $\beta_2 \gg |\beta_3|$.

*SPM effects on third-order and higher spectral phase*

The relation for a parabolic profile, $\beta_2 \gg |\beta_3|$ , is achieved via the effect of SPM on $\beta_3$. $\beta_3$'s dependency, $|\beta_3| \propto \frac{1}{\Delta\nu^2}$, on SPM bandwidth generation scales much higher than that of $\beta_2$ ($\beta_2 \propto \frac{1}{\Delta\nu}$). The higher dependence of $\beta_3$ on the bandwidth results in the peak magnitude being less than that of $\beta_2$ and the decrease of $\beta_3(\Delta z)$ being sharper versus propagation. It can be shown rearranging Eq. 2, for $\beta_3$, that $\beta_3$ scales at most with the bandwidth like[1],

$$|\beta_3(\Delta z)| \propto \frac{\tau(\Delta z))}{\Delta\nu(\Delta z)^2}, \tag{5}$$

We plot the ratio of $\beta_3$, to $\beta_{3o} = d_3 \Delta z$ along the fiber length, $d_3$ is the waveguide third order dispersion coefficient in Fig. 3c, obtained from the GNLSE numerical simulation. The

---

[1] An adjustment to the proof shown in Eq. 3 and 4 must be made for odd-order coefficients (e.g., 3rd order) because they follow the same sign as the same-order waveguide GVD coefficient and could be negative (such as the numerical example of Fig. 1). This means that the summation term of Eq. 3 is not smaller than $\tau$. However, For odd-orders, it can still be proven that $|\beta_n(\Delta z)| \propto \frac{\tau(\Delta z)}{\Delta\nu(\Delta z)^{n-1}}$ by the fact that odd orders impact the duration asymmetrically on both sides of the pulse in time. It can be shown that, instead, $|\beta_n(\Delta z)| \propto \frac{\tau_n(\Delta z)}{\Delta\nu(\Delta z)^{n-1}}$, where $\tau_n(\Delta z) = \big|\,|\tau_{1/e}| - |\tau_{-1/e}|\,\big|$. However, $\big|\,|\tau_{1/e}| - |\tau_{-1/e}|\,\big| \leq \tau = \tau_{1/e} + |\tau_{-1/e}|$. Therefore, odd-coefficients scale still in a maximal manner as $|\beta_n(\Delta z)| \propto \frac{\tau(\Delta z)}{\Delta\nu(\Delta z)^{n-1}}$.

For even-order coefficients, they are always positive because SPM generates bluer frequencies on the trailing side of the pulse and redder frequencies on the leading side of the pulse. Since the pulse is spreading in time due to the dominant ND $\beta_2$, this enforces that all even higher-order coefficients are positive.

figure shows that indeed the peak reduction and the steeper decrease of $\beta_3$ occurs in comparison to the propagation dynamics of the second-order coefficient.

Fig. 3c shows that the ratio of $\beta_3$ to $\beta_{3o}$ grows at first, to about 90 percent and then descends to a minimum of 13% at about 20% $L_{sat}$ (5 cm). In the region where bandwidth saturation starts to take place, the ratio then grows again and eventually approaches one slowly, in the asymptotical limit, for the same reasons given for the second-order coefficient case. However, even at the saturation length, the ratio is still only 74% and across the full propagation remains below one, in stark contrast to the second-order coefficient curve. Therefore, everywhere in the ND fiber, SPM substantially reduces the third-order spectral phase coefficient below the value only considering dispersion without SPM.

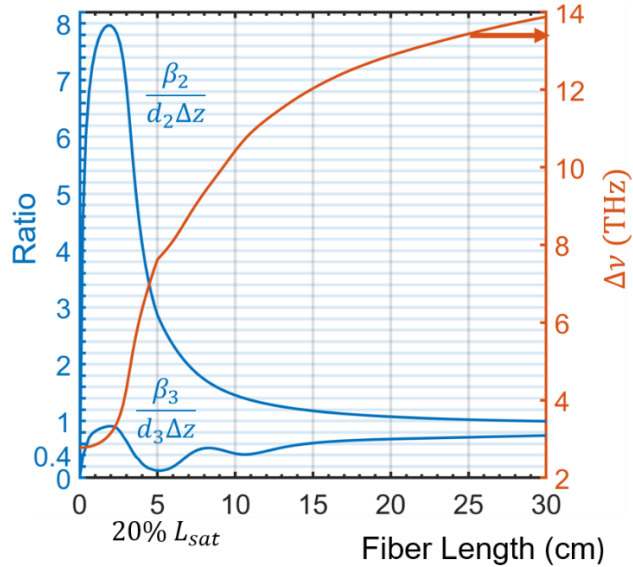

Fig. 3: a) The 1/e angular frequency bandwidth of the spectral energy density (in Hz) versus propagation distance up to the saturation length in the normal dispersion fiber. b) Plot of the ratio of the second-order spectral phase coefficient, $\beta_2$, of the SCG pulse to that when there is no nonlinear effect, $(d_2\Delta z)$, versus propagation in the normal dispersion fiber. c) Plot of the ratio of the third-order spectral phase coefficient, $\beta_3$, of the SCG pulse to the third-order spectral phase coefficient in the absence of any nonlinear effect $(d_3\Delta z)$ versus the propagation in the normal dispersion fiber. Results are obtained from the GNLSE simulation.

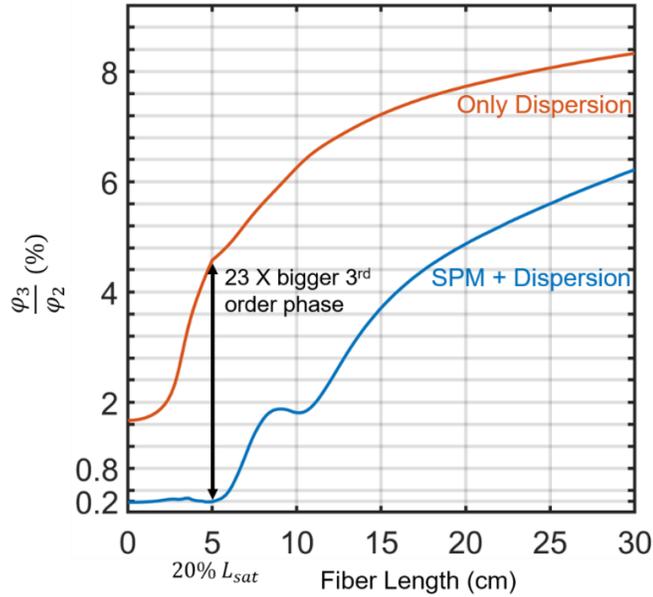

Fig. 4: Plot of the percentage of third order spectral phase contribution to second order contribution at the 1/e bandwidth of the supercontinuum pulse versus propagation distance in fiber. Results are obtained from the GNLSE simulation. Red (top curve) corresponds to case where only dispersion is present, blue (lower curve) corresponds to case with both SPM and dispersion. At 20 % the saturation length, SPM lowers the percentage of third order spectral phase to second by a factor of 23, i.e., the pulse is substantially more parabolic than with only dispersion.

The steeper decrease of $\beta_3(\Delta z)$, leads to the ratio of $\left|\frac{\beta_3(\Delta z)}{\beta_2(\Delta z)}\right|$ approximately scaling as $\left|\frac{\beta_3(\Delta z)}{\beta_2(\Delta z)}\right| \propto \frac{1}{\Delta\nu(\Delta z)}$. This relation of the two coefficients is obtained from Eq. 4, 5. The inverse bandwidth dependence of the phase ratios proves, analytically, the main claim of this section, that the phase converges to a near parabolic profile even in the presence of substantial higher order dispersion.

To extend the analysis to higher-orders, using a similar method as for the third-order coefficient, it can be shown that, maximally, $|\beta_4(\Delta z)| \propto \frac{\tau(\Delta z)}{\Delta\nu(\Delta z)^3}$. In general, maximally, $|\beta_n(\Delta z)| \propto \frac{\tau(\Delta z)}{\Delta\nu(\Delta z)^{n-1}}$, and the reductive effect of SPM on higher-order spectral phase coefficients is even more pronounced as they scale inverse exponentially to the frequency bandwidth.

### SPM leads to a parabolic phase convergence

The main claim, that the spectral phase converges to a parabolic profile and is robust to higher-order dispersion is verified numerically, from the GNLSE simulation, and shown in Fig. 4. Fig. 4 is the plot of the percentage of third-order spectral phase contribution to second-order contribution at the 1/e bandwidth frequency value versus propagation distance in the fiber. The percentage value is shown for the 1/e bandwidth frequency as here is where it is maximized within the bandwidth range, representing the maximal deviation from a parabolic phase profile.

The lower curve, of Fig. 4, corresponds to the percentage of third-order spectral phase to second-order at the 1/e bandwidth of the pulse (i.e., $100\% \times \left|\frac{\Delta\nu}{6}\frac{\beta_3}{\beta_2}\right|$) with SPM and dispersion, while the top curve corresponds to only when dispersion is present (i.e., $100\% \times \left|\frac{\Delta\nu}{6}\frac{d_3}{d_2}\right|$). The ratio with SPM contributions stays well below the same ratio where no SPM is present. This ratio dips to a factor of 23 less than the linear case (0.2%), indicating a strong convergence to a parabolic profile at 20% the spectral saturation length (5 cm- also where $\beta_3$ shows a minimum in Fig. 3c). Moreover, the ratio stays below the dispersion only case by more than a percent (corresponding to approx. 75% the value of the dispersion only case) up to the saturation length.

Fig. 4 also indicates that the percentage value at the 1/e bandwidth corresponding to the input pulse, for the dispersion only case (1.6%), equates to that of the SPM case at 3 times the original bandwidth (8 cm of length). Thus, the spectral phase is stretched by a factor of three horizontally before its values start to increase past the maximal value found only when considering dispersion (linear pulse propagation), as illustratively shown in Fig. 1.

Furthermore, the SCG bandwidth is about a factor of 2.5 greater than the input bandwidth at the propagation coordinate location where strong convergence of spectral phase to a parabolic profile occurs ( $20\% L_{sat}$ ). Both a near parabolic spectral phase profile with substantial bandwidth development, make 20% the saturation length, the ideal ND segment length, for this example, when SCG is in the SPM dominated regime in the alternating waveguide.

In general, ND segments lengths should be found by finding where strong parabolic phase convergence occurs, while spectral generation remains significant. As such, high pulse compression ratios can occur in the sign-alternating dispersion waveguide while minimizing the amount of ND segments used to obtain the compression factor.

Turning to the dispersion profile of the subsequent AD segment needed to compress the pulse coming from the ND segment, we find that the AD GVD profile must satisfy $\frac{\beta_3}{\beta_2} = \frac{d_{3,AD}}{d_{2,AD}}$ for transform limited pulse compression. Therefore, the AD GVD profile must be flat in the SPM dominated regime of SCG.

A non-ideal AD GVD profile in the SPM dominated regime of SCG, may not be a problematic criterion since the waveguide consists of subsequent periods of ND-AD segments. After the pulse emerges from this (non-ideal) AD segment, it would still go into a subsequent ND segment, where the nonlinear generation there would reduce the higher-order spectral phase variation the entering pulse has.

For example, this reduction of input phase explains the experimental result of [17]. Specifically, the experimental result shows that the maximum 3rd order phase contribution, occurring at the endpoints of 1/e bandwidth of the SCG pulse is a factor of three less than the linear case at the output. The factor of three reduction is despite uncompensated higher-order dispersion in the AD segments.

Equations 4,5 are valid for AD SCG as well. Thus, in AD SCG the spectral phase coefficients are reduced when bandwidth generation occurs rapidly compared to temporal effects of dispersion. However, the reduction is not as substantial as in ND SCG because of complex spectral phase from SPM in the wings of the pulse (no wave-breaking effect in AD).

The convergence of spectral phase to a parabolic profile regardless of higher-order dispersion has particular advantages for sign-alternating waveguides. Specifically, the choice of ND segments for pulse compression when SPM dominates is not limited by the specific shape of the dispersion profile for the ND segment.

## 5. Additional Higher-Order Phase Reduction Across Alternating Waveguides

In the context of the sign-alternating dispersion waveguides the bandwidth entering subsequent ND segments continually increases, resulting in a larger decrease in higher order spectral phase coefficients (e.g., as seen in Eq. 5). This would in turn result in a stronger convergence to a parabolic spectral phase profile versus a uniform ND SCG waveguide, across the sign-alternating dispersion structure for segments where SCG is in the SPM dominated regime. The advantage of sign-alternating dispersion waveguides for pulse compression then lies both in spectral bandwidth versus input peak power efficiency and in the spectral phase profile for pulse compression applications.

## 6. Designing Practically Feasible AD GVD Profiles for General Spectral Phase Profiles

The above analysis for the SPM dominated regime of SCG guides the choice to flat AD GVD profiles. However, obtaining these flat profiles for an ever-increasing frequency range across the sign-alternating structure is rather difficult.

Moreover, there is a transition region from SPM dominated to dispersion dominated for the ND segments in sign-alternating dispersion waveguides. For these middle ND segments in the alternating structure, the ideal AD profiles become unclear. It is then needed to derive a practical and general method for obtaining any needed AD GVD such that the pulse obtains a transform-limited duration at the end of the segment.

We proceed by splitting the AD segment into several sub-segments, for the goal of compensating the spectral phase of the pulse coming out of the ND segment. The overall GVD coefficient of these sub-segments is anomalous to compensate for the positive second-order spectral phase coefficient of the ND segment SCG. The number of such sub-segments, labeled as $n$, determines the order precision of spectral phase compensation. For example, two sub-segments would compensate up to the third-order spectral phase; three would compensate up to fourth-order and so forth. We show this AD segment scheme in Fig. 5.

Next, for an optical pulse having traversed these sub-segments, perfect spectral phase compensation up to order "$n + 1$" would be achieved if the segments satisfy:

$$-\sum_{k=1}^{n} d_{2,k} L_k = \beta_2 \equiv \frac{d^2\varphi}{d\nu^2}\Big|_{\nu_o}$$

$$-\sum_{k=1}^{n} d_{3,k} L_k = \beta_3 \equiv \frac{d^3\varphi}{d\nu^3}\Big|_{\nu_o} \tag{6a}$$

$$\vdots$$

$$-\sum_{k=1}^{n} d_{n,k} L_k = \beta_n \equiv \frac{d^{n+1}\varphi}{d\nu^{n+1}}\Big|_{\nu_o}$$

Where, $d_{j,k}$ is the $j^{\text{th}}$ dispersion coefficient (taken about $\nu_o$) of sub-segment indexed by the number $k$. $L_k$ is the length of the specific sub-segment.

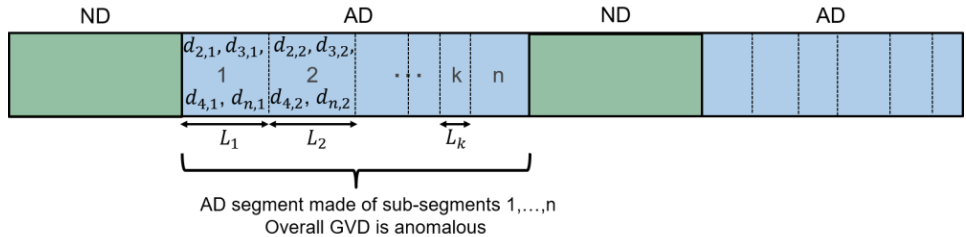

Fig. 5: AD segments can be split into subsegments where the sub-segment dispersion coefficients satisfy Eq. 6.

We represent Equation 4a in the more convenient matrix form,

$$\mathbf{Ax} = \mathbf{y} \tag{6b}$$

Where, $\mathbf{A}$ is a $n$ by $k$ matrix formed with the $d$'s as elements, i.e., $d_{j,k}$ (rows labeled by index $j$, columns by index $k$). $\mathbf{x}$ is a column vector of segment lengths, and $\mathbf{y}$ is the column vector of increasing the spectral phase derivatives of the pulse exiting the ND segment about $v_o$. The physical constraint of the above system of equations is that the segment lengths must all be positive. We also note that normal dispersion segments could be used, since the sign of higher-order dispersion coefficients are not constrained to a particular choice (ND or AD) and thus they can be used for these higher-order $d_{j,k}$ in the above system of equations. A set of unique waveguide GVD's to enable the solution of Eq. 6 could be found by varying the sub-segment geometrical cross section (e.g. such as core radius in fiber).

## 7. AD GVD in the Dispersion Dominated Regime

For the dispersion dominated regime, again, an entirely different AD segment GVD is found compared to that of the SPM dominated regime. Here, the AD GVD profile would converge to a scaled reflection of the previous ND segment GVD curve about the frequency axis, with the scaling factor determining the AD segment's length. I.e.,

$$\text{AD GVD } (v) = -c\text{ND GVD } (v), c > 0 \qquad , \qquad \text{Eq.7}$$

and the AD segment length is the previous ND segment length divided by c. This AD GVD profile emerges because the temporal dynamics become dominated by linear dispersion, where the spectral generation per ND segment is negligible compared to the input bandwidth, in accordance to what is expected in the dispersion dominated regime of SCG.

From Eq. 7 it can be seen that for a constant ND segment length, the AD segments converge to a constant length as well (provided c stays the same, i.e., the same AD and ND material is used). Then, periodic waveguides can be constructed, where the spectrum will linearly increase across the waveguide's ND segments [17], provided losses are negligible. Ultimately, this allows for sign-alternation to be used in a resonator or pulse circulator configuration.

Another advantage is that the dynamics within the dispersion dominated regime, provides another option for the design of AD segment dispersion in sign-alternating dispersion waveguides. Instead of maximal pulse compression within a few segments, where the AD segment profiles become hard to engineer, and are changing, one can use a simple periodic arrangement of highly dispersive unchanging segments. This simple arrangement would achieve the same spectral frequency bandwidth and compression ratio albeit with many more segments.

The increased number of segments make it critical to minimize and to understand how losses effect bandwidth generation, when using the dispersion dominated regime for nonlinear pulse compression. The results of [17] for the dispersion dominated regime can be extended to incorporate the effects of losses, i.e., the maximum bandwidth increase as a function of the amount of bandwidth generating segments, $n$, with losses is given as,

$$\left(\frac{1-\varepsilon^n}{1-\varepsilon}\right)\delta v, \qquad \text{Eq. 8}$$

Where, $\varepsilon$ is the segment specific loss and $\delta v \approx \left[0.81\gamma\frac{E}{4|\beta_2|}\right]$ ($E$ being the pulse energy).

## 8. Outlook towards single-cycle compression in integrated photonics

In a practical setting there are many other considerations for high-quality pulse compression in sign-alternating structures, that have not been subjects of rigorous discussion in this paper. For example, the impact of deviations of segment lengths from calculated values, and the constraints outlined in the general considerations section of the paper.

However, a more impactful concern would be that spectral generation in AD waveguide segments takes place. For example, if the input pulse energy largely exceeds the fundamental soliton energy, $E = \beta_2/(\gamma\tau_o)$ in the AD segments, where $\tau_o$ is the transform limited duration of the pulse entering the segment, substantial nonlinear compression and generation can take place in these segments even with low nonlinear coefficient.

It is then warranted to address this concern by considering the impact of spectral generation in the AD segments. We proceed by designing and simulating the nonlinear pulse evolution of a sign-alternating dispersion waveguide in a Silicon-Nitride (SiN) integrated photonic setting. We use Eq. 1 as the GNLSE of the system, where we omit self-steepening and Raman contributions. The lengths and nonlinear coefficients of each waveguide segment are summarized in table 1, while the GVD profile versus pulse envelope angular frequency is given in Fig. 5.

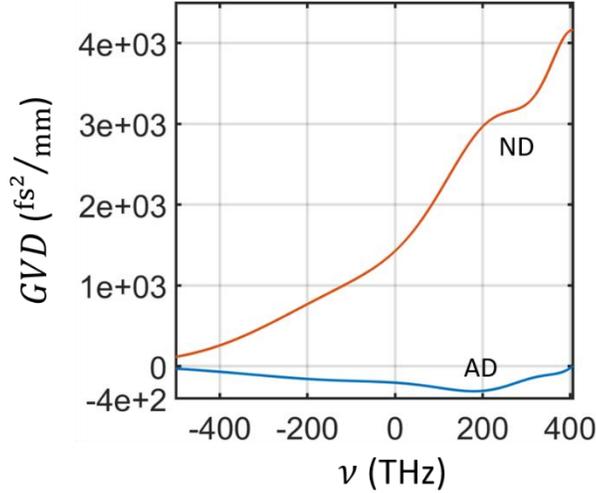

Fig 6: Group velocity dispersion versus angular frequency of the pulse envelope for both the ND segments (red) and AD segments (blue) of the Silicon Nitride sign-alternating dispersion waveguide.

The example structure we simulate was designed to show the relevant effects of nonlinear pulse compression when AD SCG dominates and to show that large pulse compression factors can be obtained, under the high pulse energy condition as well. We foresee that further optimization, especially in segment dispersion profiles and factoring other practical considerations such as losses, and manufactured length variances, can achieve pulse compression ratios much higher than shown here.

The input pulse to the constructed sign-alternating dispersion SiN waveguide has a 1/e duration of 144 fs with Gaussian profile centered at 1550 nm. The input pulse energy of 100 pJ, is substantially higher than the soliton energy in the first AD segment (approx. 6 pJ). Thus, generation is carried out in both ND and AD segment types.

The constructed sign-alternating waveguide's temporal development is indicated in Fig. 6a, where a final 1/e pulse duration of approx. 11 fs is obtained, which is approx. two optical cycles. This sign-alternated dispersion waveguide then has a temporal compression factor of 12. Also indicated in the figure are the lengths of the waveguide segments, where red indicates an ND segment and yellow an AD segment. Fig. 6b, compares the pulse power profile versus pulse time, normalized to the peak power, of both the transform limited profile and the output profile from the sign-alternating structure.

As seen in Fig. 6b, the output pulse exhibits more features below the 20 percent level of the peak power than the transform limited profile, of duration 10 fs. These pulse features are because of uncompensated spectral phase, a substantial amount of which, is from SPM in the AD segments, where the optical wave-breaking effect is not present.

|  | ND 1 | AD 1 | ND 2 | AD 2 | ND3 | AD3 |
|---|---|---|---|---|---|---|
| Length (mm) | 1.5 | 12 | 0.4 | 2.5 | 0.3 | 1.3 |
| $\gamma$ (W·m$^{-1}$) | 2.24 | 0.33 | 2.24 | 0.33 | 2.24 | 0.33 |

Table 1: Respective lengths of the ND and AD segments in the sign-alternating dispersion Silicon Nitride waveguide. Nonlinear coefficients are also given.

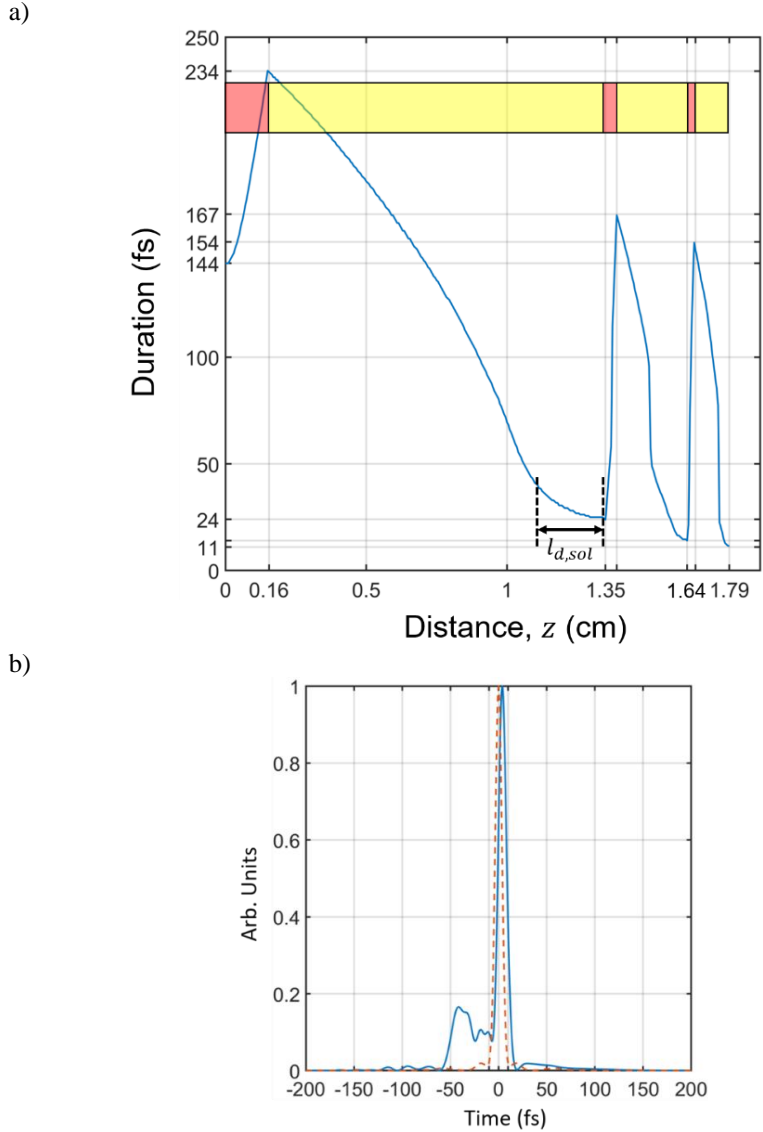

Fig. 7: A) The temporal 1/e duration as a function of distance within the sign-alternating waveguide. An output duration of 11 fs is reached, yielding a compression factor of 12. Indicated is the dispersion length of the soliton, i.e., $\frac{\tau_{sol}^2}{\beta_2}$ in the first AD segment. B) The output pulse

power profile normalized to the peak power of both the transform limited pulse (red dotted curve) associated with the spectrum and the obtained pulse from the simulation (blue curve). The power profiles are plotted against envelope time. A transform limited duration of 10 fs is obtained, while a duration of 11 fs is obtained from the simulation. The chip segments are shown in the illustration at the top (red for ND segments, yellow for AD segments).

## Design Recipe for optimum pulse compression and dynamics

We now describe in more detail the procedure we use to build our example SiN alternating waveguide, for high-energy nonlinear pulse compression and the spectral dynamics across the structure to explain the above compression factor. The first ND segment's length is chosen such that the pulse bandwidth is increased while adding minimal higher-order spectral phase. In our example, the spectral generation in this segment is kept at a minimum, so as to highlight the full development in the following AD segment. Fig. 8 shows the spectral propagation across the sign-alternating structure.

The temporally chirped pulse from the first ND segment enters a next AD segment, where, the pulse temporally compresses. Within the AD segment, the spectral dynamics can be divided into two regions as shown in Fig. 7.

The first region of AD propagation is so-named "chirped-pulse temporal compression". In this region, the pulse exhibits second order spectral phase and a stretched duration, from when it was inputted, that reduces as the pulse duration decreases over the propagation. Since the pulse reduces in duration, its peak power is raised and SPM increases its bandwidth. The bandwidth generation continually reduces the dispersion length associated with the pulse's transform limited duration across the propagation distance. The reduction in dispersion length shortens the length needed to compress the pulse to its transform limit, and ultimately the total bandwidth generation in this region.

When the pulse's spectral phase profile has zero second-order phase, nonlinear pulse compression to a soliton extends the spectral generation further, which can be substantially more than the chirped-pulse compression region (e,g, first AD segment shown in Fig. 8). This is the second region of propagation named "further soliton nonlinear compression". Within this region, the second-order phase remains zero, and the typical AD SCG dynamics of nonlinear pulse compression to the soliton fission propagation point, $z_s$ [1] occur. Namely, here to, the pulse's duration decreases further, raising peak power and SPM bandwidth generation, however, due to the pulse's high peak powers in this region, the rate of generation can be greater than in the first region, versus propagation provided its generation length is large.

While, spectral generation can be heightened within the second region in the AD segment, higher than second-order spectral phase contributions increase along with spectral modulations too. These higher-order effects are particularly dominant at the end of the region , when the pulse is within the dispersion length corresponding to the soliton that is formed i.e., when the propagation distance, $z \geq z_s - l_{d,soliton}$ , marked in Fig. 7a and 8. Here, the spectral and temporal bandwidth and duration are only slowly changing, due to the convergence to the stable soliton profile, leaving only the development of higher-order effects. The AD segment length is then chosen so that the pulse exits before it travels through this region of spectral development.

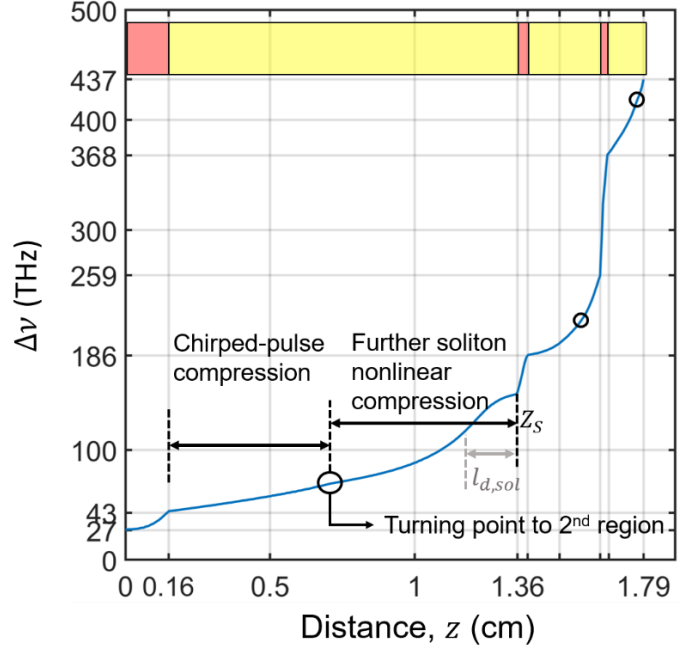

Fig 8: Spectral bandwidth development plotted against distance within the sign-alternating dispersion waveguide. The two regions of spectral generation are shown for the first AD segment and the boundaries are marked for the additional AD segments. An output spectral bandwidth of 437 THz was obtained. The chip segments are shown in the illustration at the top (red for ND segments, yellow for AD segments).

We also note, that the minimal AD segment length that can be chosen corresponds to when the second-order spectral phase becomes zero. The minimization is necessary, as a pulse with second order phase entering a subsequent ND segment would temporally defocus faster versus propagation, limiting bandwidth generation and increase higher-order phase.

In both later ND and AD segments, the saturation length and $z_s$ decreases. For later ND segments, we extend the length to a greater fraction of the saturation length to allow for more spectral development (the maximal generation allowed still decreases for subsequent segments in accordance with [17]). The extension of spectral bandwidth increase, in the ND segments is to account for the decrease in total bandwidth increase possible in subsequent AD segments across the alternating waveguide, as can be seen in Fig. 8.

This reduction of possible AD bandwidth generation is because of a reduced generation length for SPM in the chirped pulse compression regime from the smaller dispersion length associated to the pulse's increased bandwidth entering these AD segments. The further nonlinear compression region also decreases at larger bandwidths, as the energy spacing of soliton solutions decreases. Thus, a smaller amount of bandwidth generation would be sufficient to shape the pulse to a stable soliton solution for each AD segment. The bandwidth then also reduces the length for both AD regions of spectral generation which decreases the overall segment lengths (as seen in Fig. 7).

When the bandwidth of the pulse exceeds, $\Delta \nu = \frac{E}{\beta_2} \gamma$, i.e., that corresponding to the fundamental soliton, in the AD segment, the further nonlinear compression to a soliton region is not present. Thereby, past this $\Delta \nu$ bandwidth generation is only limited to the chirped pulse compression regime. This limitation then further decreases associated segment lengths and bandwidth increases.

However, spectral generation still is ongoing when only the chirped pulse compression region remains. Ultimately bandwidth generation would converge to the dynamics of the dispersion dominated SCG. In both ND and AD segments, where SCG is dominated by dispersion $\delta v$ of Eq. 8 is valid. Therefore, the bandwidth increase will converge to a linear increase in the case where spectral generation occurs in both AD and ND segments as well.

In this dispersion dominated regime, an AD GVD profile satisfying Eq. 7 would be ideal. It can be seen, from Fig. 6, that within the range of interest (-200 THz to 200 THz) Eq. 7 is approx. satisfied with $c \approx 10$. Since the AD GVD profile satisfies the constrains for the dispersion dominated regime, in later segments, where dispersion starts to be more impactful, pulse compression in the AD segments are optimized by the AD GVD profile, explaining the close final compressed duration to the transform limit.

*Additional advantages of AD SCG within sign-alternating dispersion waveguides*

Having introduced the maximum potential bandwidth criterion for the nonlinear compression region, $\Delta v = \frac{E}{\beta_2} \gamma$, it is worthwhile to point out another advantage of sign-alternation when AD SCG takes place. Without sign-alternation, i.e., in only one AD waveguide, the chirped pulse regime is not present , or only along the beginning region of the waveguide if the input pulse is chirped. In this non-alternating case, the pulse would continuously shape into a higher order soliton and then undergo soliton fission, stopping its spectral development before the fundamental soliton bandwidth can be reached. Therefore, access to the full potential bandwidth is not possible.

For example, in the first AD segment of Fig. 7 the pulse spectral development saturates as it is shaped into a higher-order soliton of bandwidth less than the fundamental soliton bandwidth. This effect is known in literature, where a (positively) chirped pulse entering an AD waveguide yields a larger spectral output bandwidth [25].

On the other hand, in sign-alternation, the nonlinear shaping is disrupted by the temporal expansion and chirp that develops in ND segments. The chirped regions then always present in the AD segments, gives way to deeper access into the nonlinear compression region, ultimately shaping the pulse to the fundamental soliton bandwidth and even beyond, over many AD segments.

For this design strategy to work, precise segment length control, over a large range of values must be practically allowable, as seen in the length scale variation in table 1. For this reason, for high pulse compression ratios, the integrated photonics platform is an ideal venue and we seek to verify our theoretical findings and our design recipe in future experiments.

## 9.  Summary

We explored the spectral phase development of the SCG, where we find a previously unknown phase effect in supercontinuum generation within normal dispersion waveguides. Namely, that the spectral phase demonstrates robustness to higher-order dispersion. Therefore, in this regime of SCG the ND segments in sign-alternating dispersion waveguides do not have be chosen for a specific dispersion profile shape.

This robustness to ND higher order dispersion is amplified within our sign-alternating structures, rendering that our waveguide concept combines a near parabolic spectral phase profile with an increased input power to bandwidth efficiency. Thus, we foresee that our scheme has the potential for few-cycle pulse compression without the onus of high peak power drive lasers, which then provides an efficient scheme for nonlinear pulse compression for a greater plethora of optical platforms such as for high repetition rate lasers and integrated photonic applications.

We then outlined a practical design method to obtain ideal AD segment GVD profiles needed to compensate all spectral phase terms coming from the corresponding ND segments. These results are not only applicable to repeated sign-alternating SCG waveguides but for general nonlinear pulse compression schemes.

The conditions that lead to the AD and ND segment lengths to converge to a constant for both segment types were explored. This has ramifications both in the practical design of these structures for SCG and nonlinear pulse compression and whether they can be used in resonator configurations.

We concluded with indicating how pulse compression works in the high-pulse energy case, specifically in the integrated photonics setting. A rigorous design strategy was described in the design of optimal pulse compression sign-alternating dispersion waveguides in this context, with an example simulation of a structure shown. The final compressed pulse had a duration of 11 fs, being two-optical cycles.

## Funding

The authors would like to acknowledge funding from the MESA+ Institute of Nanotechnology within the grant "Ultrafast switching of higher-dimensional information in silicon nanostructures".

## Acknowledgments

The authors would like to thank Klaus-Jochen Boller for useful discussions and input. Portions of this work will be presented at the High-brightness Sources and Light-driven Interactions Congress in 2020.

## Disclosures

The author is a co-inventor for a pending filed patent owned by University of Twente of which a part is related to this work. The patent application no. is 16/204,614.